\documentclass{rspublic}

\usepackage{graphicx,graphics,psfrag,epsfig}
\usepackage{color,array,tabularx}
\usepackage{hyperref}
\usepackage{amsmath,amssymb,slashed}
\usepackage{graphicx,graphics}
\usepackage{bbm,bm}
\usepackage{calc} 
\usepackage[reals]{layout} 
\usepackage{geometry}
\usepackage{setspace}
\usepackage[english]{babel}
\usepackage[ansinew,latin1]{inputenc}
\usepackage{dsfont}
\usepackage{amsthm,amsfonts,caption}

\newcommand{\Eqref}[1]{Eq.~\eqref{#1}}
\newcommand{\pat}{\partial_t}

\begin{document}

\title[Functional renormalization for the BCS-BEC crossover]{Functional renormalization for the BCS-BEC crossover}

\author[M. M. Scherer, S. Floerchinger and H. Gies]{Michael M. Scherer${}^{1}$, Stefan Floerchinger${}^{2}$ and Holger Gies${}^{1}$}

\affiliation{${}^{1}$Theoretisch-Physikalisches Institut, Max-Wien-Platz 1, D-07749
  Jena, FSU Jena, Germany\\
  ${}^{2}$Institut f{\"u}r Theoretische Physik, Universit{\"a}t Heidelberg,
Philosophenweg 16, D-69120 Heidelberg, Germany}

\label{firstpage}

\maketitle

\begin{abstract}{
      functional RG, ultracold fermionic atoms,
    BCS-BEC crossover
}
    We review the functional renormalization group (RG) approach to the
    BCS-BEC crossover for an ultracold gas of fermionic atoms. Formulated in
    terms of a scale-dependent effective action, the functional RG
    interpolates continuously between the atomic or molecular microphysics and
    the macroscopic physics on large length scales. We concentrate on the
    discussion of the phase diagram as a function of the scattering length and
    the temperature which is a paradigm example for the non-perturbative power
    of the functional RG. A systematic derivative expansion provides for both
    a description of the many-body physics and its expected universal features
    as well as an accurate account of the few-body physics and
    the associated BEC and BCS limits.
\end{abstract}

\section{Introduction}

Many challenges in contemporary theoretical physics deal with strongly
interacting quantum field theories or many-body systems. Progress often relies
on the construction of exact or approximate solutions. In the absence of exact
solutions, reliable and controlled approximation methods typically are the
only source of information about the system and the underlying physical
mechanisms. An approximation scheme may be considered as reliable and
controlled if it is based on a systematic and consistent expansion scheme and
shows convergence towards the exact result (which, however, is often not
known). Textbook examples are, of course, provided by perturbative expansions
or lattice discretizations both of which can be consistently evaluated to a
given order or lattice refinement, systematically improved, and the
convergence can at least be checked as a matter of practice.

In this contribution, we would like to demonstrate that the functional RG can
be used to develop systematic and consistent expansion schemes for strongly
interacting systems.  Most importantly, it can be applied in the spacetime
continuum and does not require a perturbative ordering scheme. Nevertheless, it
offers a variety of tools to verify qualitative and quantitative reliability
and practical convergence. As a prime example of strongly interacting
many-body systems, we take the BCS-BEC crossover as an illustration for the use
of the functional RG. The concrete physical system that we have in mind is an
ultracold atomic Fermi gas with two accessible hyperfine spin states near a
Feshbach resonance, showing a smooth crossover between Bardeen-Cooper-Schrieffer
(BCS) superfluidity and Bose-Einstein condensation (BEC) of diatomic molecules
(Eagles 1969; Leggett 1980).

\begin{figure}[b!]
\centering
 \includegraphics[width=0.35\textwidth]{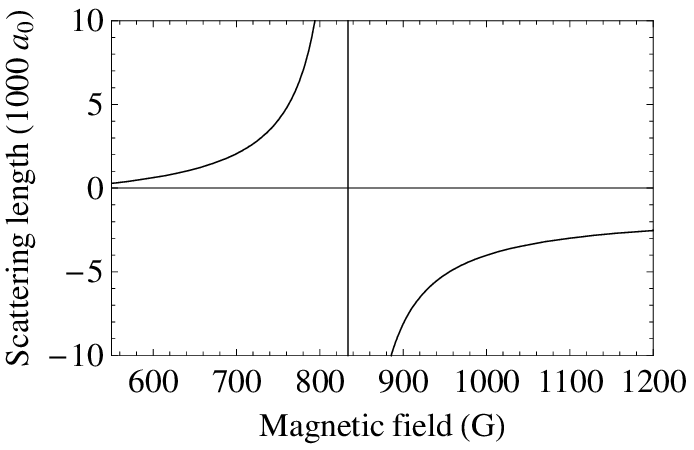}\hspace{0.2cm}
  \includegraphics[width=0.6\textwidth]{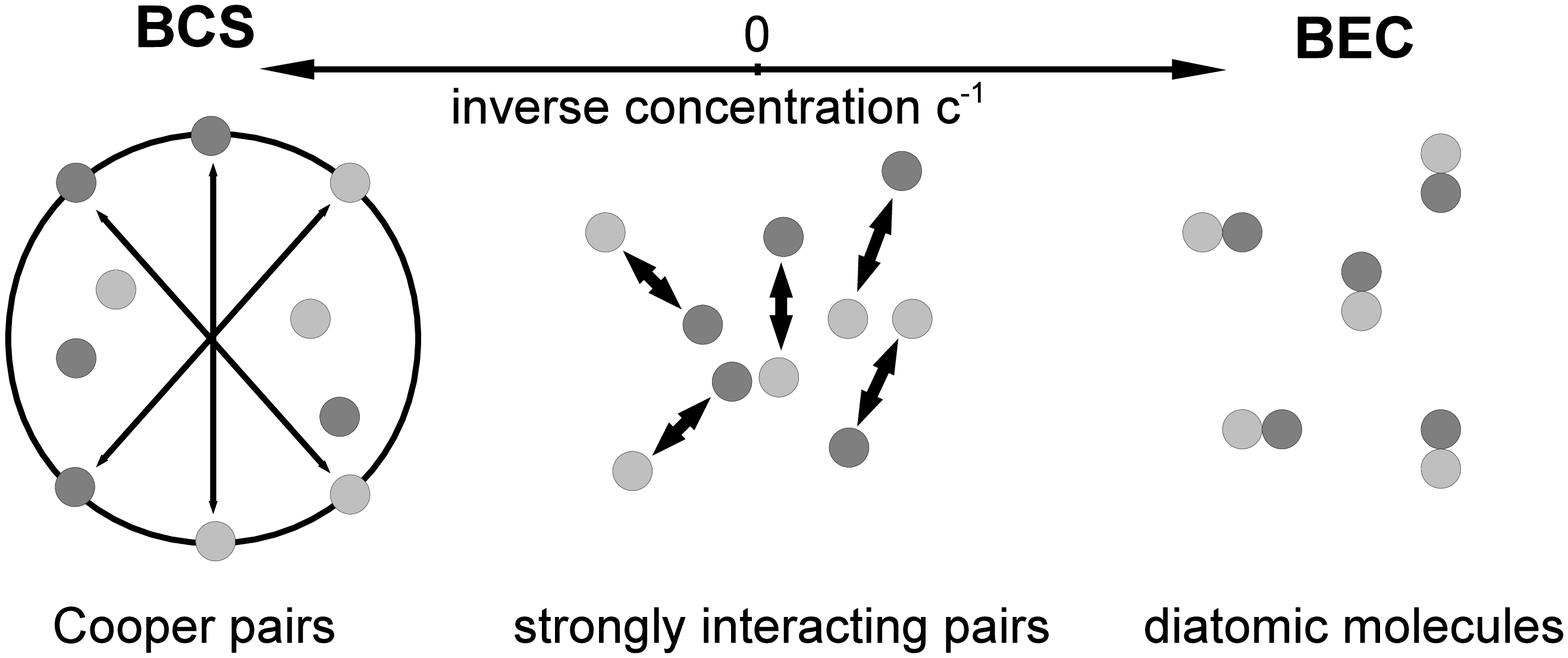}
\caption{Left: Schematic plot of a Feshbach resonance, see e.g. O'Hara (2002). Right: Sketch of
  the crossover physics.}
\label{fig:Feshbach}
\end{figure}

By means of an external magnetic field $B$ the phenomenon of a Feshbach
resonance allows to arbitrarily regulate the effective interaction strength of
the atoms, parameterized by the s-wave scattering length $a$. We briefly
discuss the example of ${}^6\mathrm{Li}$ (O'Hara \emph{et al.} 2002), which is
besides ${}^{40}\mathrm{K}$ realized in current experiments
(Regal \emph{et al.} 2004; Zwierlein \emph{et al.} 2004; Kinast \emph{et al.} 2004; Bourdel \emph{et al.} 2004; Bartenstein \emph{et al.} 2004; Partridge \emph{et al.} 2005),
see left panel of Fig. \ref{fig:Feshbach}.

For magnetic fields larger than $\sim 1200 \mathrm{G}$ the scattering length
$a$ is small and negative, giving rise to the many-body effect of
Cooper-pairing and a BCS-type ground state below a critical temperature. The
BCS ground state is superfluid described by a non-vanishing order parameter
$\phi_0 = \langle \psi_1\psi_2\rangle$ bilinear in the fermion fields. An
increase of the temperature leads to a second order phase transition to a
normal fluid, $\phi_0 =0$.  Magnetic fields below $B\sim 600\mathrm{G}$ induce
a small and positive scattering length $a$ and the formation of a diatomic
bound state, a dimer. The ground state is a BEC of repulsive dimers and again
a phase transition from a superfluid, $\phi_0 >0$, to a normal fluid, $\phi_0
=0$, can be observed at a critical temperature.  For magnetic fields in the
regime $700 \mathrm{G}\lesssim B \lesssim 1100 \mathrm{G}$ the modulus of the
scattering length $|a|$ is large and diverges at the \emph{unitary point},
$B_0=834\mathrm{G}$ where unitarity of the scattering matrix solely
  determines the two-body scattering properties.  At and near unitarity the
fermions are in a strongly interacting regime. It connects the limits of BCS
superfluidity and Bose-Einstein condensation by a continuous crossover and
also shows a superfluid ground state, with $\phi_0>0$ (Eagles 1969; Leggett
1980).

A convenient parameterization of the crossover is given by the inverse
concentration $c^{-1}=(ak_F)^{-1}$. Here the density of atoms
$n=k_F^3/(3\pi^2)$ defines the Fermi momentum $k_F$ in natural units with
$\hbar=k_B=2M=1$, where $M$ is the mass of the atoms. The inverse concentration
$c^{-1}$ varies from large negative values on the BCS side to large positive
values on the BEC side with a zero-crossing at the unitary point, see right
panel of Fig. \ref{fig:Feshbach}.

A description of the qualitative features of the BCS-BEC crossover has been achieved by Nozieres \& Schmitt-Rink (1985) and Sa de Melo \emph{et al.} (1993) within extended mean-field theories which account for the contribution of both fermionic and bosonic degrees of freedom.
 However, the quantitatively precise understanding of BCS-BEC
crossover physics requires non-perturbative methods. The experimental
realization of molecule condensates and the subsequent crossover to a BCS-like
state of weakly attractive fermions
(Regal \emph{et al.} 2004; Zwierlein \emph{et al.} 2004; Kinast \emph{et al.} 2004; Bourdel \emph{et al.} 2004; Bartenstein \emph{et al.} 2004; Partridge \emph{et al.} 2005)
pave the way to future experimental precision measurements and provide a
testing ground for non-perturbative methods.  An understanding of the crossover on a quantitative level at and near the resonance has been developed through numerical
Quantum Monte-Carlo (QMC) methods
(Carlson \emph{et al.}  2003; Astrakharchik \emph{et al.} 2004; Bulgac \emph{et al.} 2006; Burovski \emph{et al.} 2006; Akkineni \emph{et al.} 2007). The complete phase diagram has been accessed by functional
field-theoretical techniques, as $t$-matrix approaches
(Pieri \& Strinati 2000; Perali \emph{et al.} 2004), Dyson-Schwinger equations
(Diehl \& Wetterich 2006, $b$), 2PI methods (Haussmann \emph{et al.} 2007), and RG flow
equations (Birse \emph{et al.} 2005; Diehl \emph{et al.} 2007$a$, $b$; Gubbels
\& Stoof 2008; Bartosch \emph{et al.} 2009; Krippa 2009). These
pictures of the whole phase diagram
(Nikolic \& Sachdev 2007; Pieri \& Strinati 2000, Perali \emph{et al.} 2004; Diehl \& Wetterich 2006, $b$; Birse \emph{et al.} 2005; Diehl \emph{et al.} 2007$a$, $b$; Gubbels \& Stoof 2008, Haussmann \emph{et al.} 2007) do not yet reach a similar
quantitative precision as anticipated for the QMC calculations.

We intend to fill this gap and discuss the limit of broad Fesh\-bach
resonances for which all thermodynamic quantities can be expressed in terms of
two dimensionless parameters,
\begin{eqnarray}
\text{the concentration:} && c=ak_F\,,
\end{eqnarray}
and the temperature in units of the Fermi temperature $T/T_F$ where
$T_F=k_F^2$. In the broad resonance regime, macroscopic observables are
\emph{universal} (Diehl \& Wetterich 2006, Nikolic \& Sachdev 2007, Diehl \emph{et al.} 2007$a$, Ho 2004),
i.e. they are to a large extent independent of the concrete microscopic
realization. Very similar to the notion of universality near 2nd-order
  phase transitions, universality in the present context can be traced back
to the existence of a fixed point in the RG 
flow which is approached provided the Feshbach (Yukawa) coupling is large enough
(Diehl \emph{et al.} 2007$b$).

 Our review is based on the RG studies of (Diehl \emph{et al.} 2007$a,b$;
  Diehl \emph{et al.} 2010; Floerchinger \emph{et al.} 2009; Floerchinger \emph{et al.} 2010). For RG studies
  using different expansion schemes, see (Birse \emph{et al.} 2005; Gubbels \&
  Stoof 2008; Bartosch \emph{et al.} 2009; Krippa 2009).  In the
  following, we first introduce the required techniques from RG flow
  equations for cold atoms, see Secs. \ref{sec:micmodBCSBEC} -
  \ref{sec:BCSBECvac}. Further, we include the quantitative effect of
  particle-hole fluctuations, Sec. \ref{sec:BCSBECph}, and systematically
  extend the truncation scheme, accounting for changes of the Fermi surface
  due to fluctuation effects, Sec. \ref{sec:BCSBECrunfermion}. Additionally,
  we also consider an atom-dimer interaction term.


\section{Microscopic model and the functional RG}\label{sec:micmodBCSBEC}
Microscopically the BCS-BEC crossover can be described by an action including
a two-component Grassmann field $\psi=(\psi_1,\psi_2)$, describing
non-relativistic fermions in two hyperfine states and a complex scalar field
$\phi$ as the bosonic degrees of freedom. In different regimes of the
crossover $\phi$ can be seen as a field describing molecules, Cooper pairs or
simply an auxiliary field.  Explicitly, the microscopic action at the
ultraviolet scale $\Lambda$ reads
\begin{equation}
S  = \int_0^{1/T} d\tau \int d^3x{\Big \{}\psi^\dagger(\partial_\tau-\Delta-\mu)\psi
 +\phi^*(\partial_\tau-\frac{\Delta}{2}-2\mu+ \nu_\Lambda(B))\phi- h_\Lambda(\phi^*\psi_1\psi_2+h.c.){\Big \}}.
\label{eqMicroscopicAction}
\end{equation}
In thermal equilibrium, the system is described by the Matsubara formalism. The variable $\mu$ is the chemical potential.  The
parameter $\nu_\Lambda(B)=\nu(B)+\delta \nu(\Lambda)$ includes the detuning from
the Feshbach resonance $\nu(B)=\mu_\mathrm{M}(B-B_0)$, with the magnetic
moment of the boson field $\mu_\mathrm{M}$ , and a renormalization counter
term $\delta \nu(\Lambda)$ that has to be adjusted to match the conditions
from the physical vacuum. The Yukawa coupling $h_\Lambda$ is related to the
width of the Feshbach resonance. Note that the bosonic field $\phi$ appears
quadratically in Eq. \eqref{eqMicroscopicAction} so the functional integral
over $\phi$ can be carried out, corresponding to an inverse
  Hubbard-Stratonovich transformation, and our model is equivalent to a purely
fermionic theory with an interaction term
\begin{equation}
S_{\mathrm{int}}  = \int_{p_1,p_2,p_1^\prime,p_2^\prime}\lambda_{\psi,\mathrm{eff}}(p_1+p_2) \psi_1^{\ast}(p_1^\prime){\psi_1}(p_1) \psi_2^{\ast}(p_2^\prime){\psi_2}(p_2)\,\delta(p_1+p_2-p_1^\prime-p_2^\prime).
\label{lambdapsieff}
\end{equation}
Here, $p = (p_0,\vec p)$ and the microscopic interaction between the fermions
is described by the tree-level expression with a classical inverse boson
propagator in the denominator
\begin{equation}\label{eq:Bosonpropagator}
\lambda_{\psi,\mathrm{eff}}(q)= -\frac{h_\Lambda^2}
{ -\omega + \frac{\vec{q}^2}{2}-2\mu+ \nu_\Lambda (B)}\,, 
\end{equation}
where $\omega$ is the real-time frequency of the exchanged boson $\phi$. It is
related to the Matsubara frequency $q_0$ via analytic continuation
$\omega=-iq_0$. Further, $\vec q=\vec p_1+\vec p_2$ is the center of mass
momentum of the scattering fermions $\psi_1$ and $\psi_2$ with momenta $\vec
p_1$ and $\vec p_2$, respectively.  The limit of broad Feshbach resonances
corresponds to $h_\Lambda\to\infty$. In this limit the microscopic interaction
becomes pointlike, with strength $-h_\Lambda^2/\nu_\Lambda$.

The functional RG can be formulated as a functional differential equation for an action
functional for which the microscopic model serves as an initial value. Whereas
the microscopic interactions are governed by $S$ at the ultraviolet scale
$\Lambda$, quantum and thermal fluctuations effectively modify the
interactions at larger length scales which can be summarized in an effective
action $\Gamma_k$ (e.g. for the 1PI proper vertices) valid at a momentum scale
$k$. In other words, $\Gamma_k$ includes the effects of fluctuations with
momenta higher than $k$ and governs the interactions with momenta near
$k$. This {\em effective average action} or {\em flowing action} satisfies the Wetterich equation
(Wetterich 1993), being an exact RG flow equation,
\begin{equation}\label{eq:flowequwett}
 \partial_k \Gamma_k[\Phi]=\frac{1}{2}\mathrm{STr}\left[ 
\left(\Gamma_k^{(2)}[\Phi]+R_k \right)^{-1} \partial_k R_k\right] \,.
\end{equation}
Here, the STr operation involves an integration over momenta and a summation
over internal indices with appropriate minus signs for fermions. The 
collective field $\Phi$ summarizes all bosonic and fermionic degrees of
freedom, and $\Gamma_k^{(2)}[\Phi]$ denotes the second functional derivative
of $\Gamma_k$
\begin{equation}
 \left( \Gamma_k^{(2)}[\Phi]\right)_{ij}(p_1,p_2)=\frac{\overrightarrow\delta}{\delta\Phi_{i}(-p_1)}\Gamma_k[\Phi]\frac{\overleftarrow\delta}{\delta\Phi_{j}(p_2)}\,.
\end{equation} 
The long-wavelength regulator $R_k$ specifies the details of the
regularization scheme. Specific examples will be discussed below.  For reviews
of the functional renormalization group see (Salmhofer \& Honerkamp 2001;
Berges \emph{et al.}  2002; Gies 2006; Pawlowski 2007; Kopietz 2010).  From the full
effective action in the long wavelength limit
$\Gamma[\Phi]=\Gamma_{k=0}[\Phi]$, all macroscopic properties of the system
under consideration can be read off.

Equation (\ref{eq:flowequwett}) is
the technical starting point of our investigations. It is a functional
differential equation which, upon expansion of this functional into a suitable
basis translates to a system of infinitely many coupled differential equations
for the expansion coefficients, i.e. generalized running couplings.
Identifying suitable expansion schemes is not a formal but a physics problem:
expansions should be based on building blocks that encode the relevant degrees
of freedom of the system possibly at all scales. In the present context, this
emphasizes the usefulness of composite bosonic fields which are expected to be
the relevant long-range degrees of freedom at low temperatures. Reducing the
full effective action to a treatable selection of generalized couplings
defines a \emph{truncation}. Possible truncation schemes include vertex
expansions, derivative expansions or other schemes to systematically classify
all possible operators of a given system. The quantitative success of a given
truncation scheme does not necessarily rely on the existence of a small
expansion parameter like the interaction strength, but only requires that the
operators neglected in a truncation do not take a strong influence on the flow
of the operators included. In practice, a truncation can be tested in various
ways, e.g., by verifying the practical convergence for increasing truncations
or by studying regulator-scheme independencies for universal quantities. In
the present context, also the comparison with well-known few-body physics
turns out to provide a useful benchmark.


\section{Basic truncation}\label{sec:BCSBECtrunc}

\subsection{Derivative expansion}
Thermodynamics of a system can be obtained from its grand canonical partition
function $Z$ or the corresponding grand canonical potential $\Omega_G=-T\,
\mathrm{ln}Z$. It is related to the effective action via $
\Gamma[\Phi_{eq}]=\Omega_G/T$, where $\Phi_{eq}$ is obtained from the field equation
$\frac{\delta}{\delta\Phi}\Gamma[\Phi] |_{\Phi=\Phi_{eq}}=0$.
Let us first present a basic version of a truncation which already captures all
the qualitative features of the BCS-BEC crossover:
\begin{equation}
\Gamma_k[\Phi]  =   \int_{\tau,\vec{x}} {\bigg \{} \psi^\dagger (\partial_\tau -\Delta -\mu) \psi+  \bar{\phi}^*\Big(\bar Z_\phi \partial_\tau-\frac{A_\phi \Delta}{2}\Big)\bar\phi + \bar U(\bar \rho,\mu) -\bar h (\bar \phi^* \psi_1\psi_2 + \bar \phi \psi_2^\ast \psi_1^\ast) {\bigg \} }.
\label{eq:baretruncation}
\end{equation}
The effective potential $\bar U(\bar \rho,\mu)$ is a function of
$\bar{\rho}=\bar{\phi}^*\bar{\phi}$ and $\mu$. This truncation can be
motivated by a systematic derivative expansion and an analysis of the
  symmetries encoded in Ward
identities (Diehl \emph{et al.} 2007$a$; Diehl \emph{et al.} 2010). It does not yet
incorporate, for instance,
the effects of particle-hole fluctuations and we will come back to this issue
in Sect. \ref{sec:BCSBECph}. We define renormalized fields $\phi=
A_\phi^{1/2}\bar\phi$, $\rho= A_\phi \bar \rho$ and renormalized couplings
$Z_\phi=\bar Z_\phi / A_\phi$, $h=\bar h/\sqrt{ A_\phi}$ and express
Eq. \eqref{eq:baretruncation} in these quantities
\begin{equation}
\Gamma_k[\Phi]  = \int_{\tau,\vec{x}} {\bigg \{}  \psi^\dagger (\partial_\tau -\Delta -\mu) \psi+  \phi^*\left( Z_\phi \partial_\tau-\frac{ \Delta}{2}\right)\phi+  U(\rho,\mu) -  h \,(\phi^* \psi_1\psi_2 + \phi \psi_2^\ast \psi_1^\ast) {\bigg \} }.
\label{eq:truncation}
\end{equation}
We expand the effective potential around the $k$-dependent
location of the minimum $\rho_0(k)$ and the $k$-independent value of the
chemical potential $\mu_0$ corresponding to the physical particle number
density $n$. We determine $\rho_0(k)$ and $\mu_0$ by the requirements
$(\partial_\rho U)(\rho_0(k),\mu_0)=0$ for all $k$, and $-(\partial_\mu U)
(\rho_0,\mu_0)=n$ at $k=0$.  More explicitly, we employ a simple
  expansion for
$U(\rho,\mu)$ of the form
\begin{equation}\label{eq:BCSBECueff}
U(\rho,\mu) = U(\rho_0,\mu_0)-n_k (\mu-\mu_0) +(m^2+\alpha(\mu-\mu_0)) (\rho-\rho_0) +\frac{1}{2}\lambda (\rho-\rho_0)^2.
\end{equation}
In the symmetric or normal gas phase, we have $\rho_0=0$, while in the phase
with spontaneous breaking of $U(1)$ symmetry (superfluid phase), we have
$\rho_0>0$ and $m^2=0$. The atom density $n=-\partial U/\partial \mu$
corresponds to $n_k$ in the limit $k\to 0$.

The running couplings in this truncation explicitly are $m^2(k)$, $\lambda(k)$, $\alpha(k)$,
$n_k$, $Z_\phi(k)$ and $h(k)$. In the phase with spontaneous symmetry breaking
$m^2$ is traded for $\rho_0$. In addition, we need the anomalous dimension
$\eta=-k\partial_k \text{ln} A_\phi$.
At the microscopic scale $k=\Lambda$ the initial values of our couplings are
determined from Eq. \eqref{eqMicroscopicAction}. This gives
$m^2(\Lambda)=\nu_\Lambda(B)-2\mu$, $\rho_0(\Lambda)=0$, $\lambda(\Lambda)=0$,
$Z_\phi(\Lambda)=1$, $h(\Lambda)=h_\Lambda$, $\alpha(\Lambda)=-2$ and
$n_\Lambda=3\pi^2 \mu\, \theta(\mu)$.
Finally, our regularization scheme is specified by a regulator for space-like momenta
which for the fermionic and bosonic field components reads
\begin{eqnarray}
\nonumber
R_{k,\psi}=\left(\mathrm{sign}(\vec p^2-\mu)k^2-(\vec p^2-\mu)\right) 
\theta\left(k^2-|\vec p^2-\mu|\right), \quad
R_{k,\phi}= A_\phi \left(k^2-\vec p^2/2\right)\theta\left(k^2-\vec
  p^2/2\right),
\label{eq:cutoff}
\end{eqnarray}
respectively. For the fermions, it regularizes fluctuations around the
Fermi surface, whereas bosonic fluctuations are suppressed for generic small
momenta. This choice is optimized in the spirit of
Litim (2000) and Pawlowski (2007).



For our choice of the regulator and with the basic approximation scheme
Eq. \eqref{eq:truncation} the flow equation for the effective potential can be
computed straightforwardly:
\begin{equation}
 k\partial_k U = \eta \rho\ U^\prime+\frac{\sqrt{2}k^5}{3\pi^2Z_\phi}\left(1-\frac{2\eta}{5}\right)s_\mathrm{B}^{(0)}-\frac{k^4}{3\pi^2}\big((\mu+k^2)^{\frac{3}{2}}\theta(\mu+k^2)-(\mu-k^2)^{\frac{3}{2}}\theta(\mu-k^2)\big)s_\mathrm{F}^{(0)},
\end{equation}
with the threshold functions
\begin{eqnarray}
s_\mathrm{B}^{(0)}=&\Big(\sqrt{\frac{k^2+U^\prime}{k^2+U^\prime+2\rho U^{\prime\prime}}}+\sqrt{\frac{k^2+U^\prime+2\rho U^{\prime\prime}}{k^2+U^\prime}}\Big)\Big(\frac{1}{2}+N_\mathrm{B}\Big[\frac{\sqrt{k^2+U^\prime}\sqrt{k^2+U^\prime+2\rho U^{\prime\prime}}}{Z_\phi}\Big]\Big),\\
s_\mathrm{F}^{(0)}=&\hspace{-6cm}\frac{2}{\sqrt{k^4+h^2\rho}}\Big(\frac{1}{2}-N_\mathrm{F}\Big[\sqrt{k^4+h^2\rho}\Big]\Big).
\end{eqnarray}
The threshold functions exhibit a temperature dependence via the Bose and
Fermi functions $N_\mathrm{B/F}[\epsilon]=(e^{\epsilon/T}\mp 1)^{-1}$. From
the effective potential flow, we derive the flow equations for the running
couplings $m^2$ or $\rho_0$ and $\lambda$. For details we refer to
(Diehl \emph{et al.} 2010). Further we need flow equations for $A_\phi$ and $Z_\phi$
that are obtained by the projections
\begin{equation}
 \pat \bar Z_\phi = -\pat \frac{\partial}{\partial q_0}(\bar P_\phi)_{12}(q_0,0)\Big|_{q_0=0},\quad \text{and}\quad \pat A_\phi = 2\pat \frac{\partial}{\partial \vec{q}^2}(\bar P_\phi)_{22}(0,\vec{q})\Big|_{\vec{q}=0},
\end{equation}
with the momentum-dependent part of the propagator 
\begin{equation}
 \frac{\delta^2 \Gamma_k}{\delta \bar \phi_i(q)\delta \bar \phi_j(q^\prime)}\Big|_{\bar\phi_1=\sqrt{2\bar\rho_0},\bar\phi_2=0}=(\bar P_\phi)_{ij}(q)\delta(q+q^\prime).
\end{equation}
Here the boson field is expressed in a basis of real fields
$\bar\phi(x)=\frac{1}{\sqrt{2}}(\bar\phi_1(x)+i\bar\phi_2(x))$. These flow
equations are derived by Diehl \emph{et al.} (2010) and have a rather involved
structure. Finally, we need the flow of the Yukawa coupling. In the symmetric
regime the loop contribution vanishes and the flow is given
by the anomalous dimension,
\begin{equation}\label{eq:BCSBECYuk}
\pat h = \frac{1}{2}\eta h,\quad \text{or in dimensionless units}\quad \pat \tilde h^2= (-1+\eta)\tilde h^2,
\end{equation}
where $\tilde h^2= h^2/k$. In the regime of spontaneous symmetry breaking ($\rho_0>0$) there
is a loop contribution $\sim h^3\lambda\rho_0$ from a mixed diagram involving both
fermions and bosons. This contribution is quantitatively subleading
which we have verified also numerically. For the basic approximation scheme,
Eq. \eqref{eq:truncation}, we therefore dropped this contribution.


\subsection{Vacuum limit and contact to experiment}\label{sec:BCSBECvac}

The vacuum limit allows us to make contact with experiment. We find that for
$n=T=0$ the crossover at finite density turns into a second-order phase
transition in vacuum (Diehl \& Wetterich 2007, Nikolic \& Sachdev 2007) as a function of the initial value
$m^2(\Lambda)$. In order to see this, we consider the momentum-independent
parts in both the fermion and the boson propagator, $-\mu$ (the ``chemical
potential'' for the fermions in vacuum) and $m(k=0)^2$, which act as gaps for
the propagation of fermions and bosons. We find the following constraints,
separating two different branches of the physical vacuum
 (Diehl \& Wetterich 2007),
\begin{eqnarray}\label{VacCond}
  \begin{array}{l l l}
    { m^2(0) >0, \quad \mu = 0  }& \text{atom phase}
      & (a^{-1} < 0) , \\
    { m^2(0) = 0,\quad \mu < 0   }& \text{molecule phase}
      & (a^{-1} > 0) ,  \\
    { m^2(0) = 0,\quad \mu = 0  }& \text{resonance}
      & (a^{-1} = 0).
  \end{array}
\end{eqnarray}
The initial values $m^2(\Lambda)$ and $h_\Lambda$ can be connected to the
two-particle scattering in vacuum close to a Feshbach resonance. For this
purpose, one follows the flow of $m^2(k)$ and $h(k)$ in vacuum, e.g. on the BCS ($a^{-1} < 0$),
i.e. $\mu=T=n=0$, and extracts the renormalized parameters $m^2=m^2(k=0)$,
$h=h(k=0)$. They have to match the physical conditions formulated in
Eq. \eqref{VacCond}. We obtain the two relations
\begin{equation}\label{eq:twobody}
\bar m^2(\Lambda)= \mu_\mathrm{M}(B-B_0)-2\mu +\frac{\bar
  h_\Lambda^2}{6\pi^2}\Lambda,\quad
 a=-\frac{h^2(k=0)}{8\pi\ m^2(k=0)}=-\frac{\bar h^2(\Lambda)}{8\pi\
   \mu_\mathrm{M}(B-B_0)},
\end{equation}
where $\mu_\mathrm{M}$ is the relative magnetic moment of the molecules.
These relations fix the initial
conditions of our model completely and similar reasoning confirms their validity on the BEC side. Now we can express the parameters
$m^2(\Lambda)$ and $h^2(\Lambda)$ by the experimentally accessible quantities
$B-B_0$ and $a$. They remain valid also for non-vanishing density and
temperature, as long as the UV cutoff $\Lambda$ is much larger than $T$ 
and $\mu$.  

\subsection{Many-body phase diagram}
Although our flow equations describe accurately the vacuum limit and can be used to determine interesting few-body parameters they are not restricted to that limit. In fact, for nonzero temperature and density, the flow deviates from its vacuum form at scales with $k^2< T$ or $k^2< T_F$. The resulting system of ordinary coupled
differential equations is then solved numerically for different chemical
potentials $\mu$ and temperatures $T$. For temperatures sufficiently small
compared to the Fermi temperature, $T/T_F\ll 1$, we find that the effective
potential $U$ at the macroscopic scale $k=0$ develops a minimum at a nonzero
field value $\rho_0>0$, $\partial_\rho U(\rho_0)=0$. The system is then in the
superfluid phase. For larger temperatures we find that the minimum is at
$\rho_0=0$ and that the ``mass parameter'' $m^2$ is positive,
$m^2=\partial_\rho U(0)>0$. The critical temperature $T_c$ of this phase
transition between the superfluid and the normal phase is then defined as the
temperature where
\begin{equation}
\rho_0=0,\quad \partial_\rho U(0)=0\quad \text{at} \quad k=0.
\end{equation}
Throughout the whole crossover the transition $\rho_0\to0$ is continuous as a
function of $T$ demonstrating that the phase transition is of second
order. An analysis of the scaling of the correlation length confirms
  that the phase transition is governed by a Wilson-Fisher fixed point for the
  $N=2$ universality class throughout the crossover (Diehl \emph{et al.}
  2010). This reflects the fact that the symmetries are properly encoded
also on the level of the truncated action.

From the flow equations together with the initial conditions we
can already recover all the qualitative features of the BCS-BEC crossover,
e.g. compute the phase diagram for the phase transition to superfluidity. The
result for this basic approximation is displayed in the right panel of Fig.\ \ref{fig:tcrit} by the dot-dashed line.

\subsection{Fixed-point and universality}

In the vacuum limit and in the regime where $k^2\gg-\mu$ the flow of the
anomalous dimension reads $\eta=h^2/(6\pi^2k)$ (Diehl \emph{et al.}
2010). Together with the dimensionless flow of the Yukawa coupling,
Eq. \eqref{eq:BCSBECYuk}, this reveals the existence of an IR attractive fixed
point given by $\eta=1,\ \tilde h^2=6\pi^2$. This fixed-point is approached
rapidly if the initial value of $h^2(\Lambda)/\Lambda$ is large enough, i.e.,
in the broad-resonance limit. Then the memory of the microscopic value of
$h^2(\Lambda)/\Lambda$ is lost at large length scales. Also all other
parameters except for the mass term $m^2$ are attracted to IR fixed points, giving
rise to universality. The fixed-point structure remains similar for
non-vanishing density and temperature and these findings also apply in this
regime and determine the critical physics of these non-relativistic quantum
fields.  For a given temperature, this fixed point has only one relevant
direction which is related to the detuning of the resonance $B-B_0$.

\section{Particle-hole fluctuations}\label{sec:BCSBECph}
\subsection{Gorkov's correction to BCS theory}

For small and negative scattering length $c^{-1}<0, |c|\ll 1$ (BCS side), the
system can be treated by the perturbative BCS theory of superfluidity
(Cooper 1956; Bardeen \emph{et al.} 1957). However, there is a significant decrease of the critical
temperature as compared to the original BCS result due to a screening effect
of particle-hole fluctuations in the medium (Gorkov \& Melik-Barkhudarov 1961; Heiselberg \emph{et al.} 2000).  Here we
will sketch the technique to include the effect of particle-hole fluctuations
in our functional RG treatment as developed by Floerchinger \emph{et al.} (2009).

\begin{figure}[b!]
\centering
\includegraphics[width=0.45\textwidth]{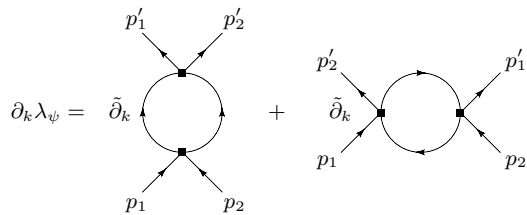}
\caption{Running of the momentum-dependent vertex $\lambda_{\psi}$. Here
  $\tilde{\partial}_k$ indicates scale-derivatives with respect to the regulator
  in the propagators but does not act on the vertices.}
\label{fig:lambdaflow}
\end{figure}

In an RG setting, the features of BCS theory can be described in a purely
fermionic language with the fermion interaction vertex $\lambda_\psi$ as the
only scale-dependent object. In general, the interaction vertex is momentum
dependent, $\lambda_{\psi}(p_1^\prime,p_1,p_2^\prime,p_2)$, and its flow has
two contributions which are depicted in Fig. \ref{fig:lambdaflow}, including
the external momentum labels. For $k \rightarrow 0,\ \mu \rightarrow 0,\ T
\rightarrow 0$ and $n \rightarrow 0$ this coupling is related to the
scattering length, $a= \frac{1}{8\pi} \lambda_{\psi}(p_i=0)$.

In the BCS approximation only the first diagram in Fig. \ref{fig:lambdaflow},
the particle-particle (pp) loop, is kept and the momentum dependence of the
four fermion coupling is neglected, by replacing
$\lambda_{\psi}(p_1^\prime,p_1,p_2^\prime,p_2)$ by the pointlike coupling
evaluated at zero momentum. For $\mu>0$, its effect increases as the
temperature $T$ is lowered. For small temperatures $T\leq T_{c,\text{BCS}}$
the logarithmic divergence leads to the appearance of pairing, as
$\lambda_\psi\to \infty$.  In terms of the scattering length $a$, Fermi
momentum $k_F$ and Fermi temperature $T_F$, the critical temperature is found
to be
\begin{equation}
T_{c,\mathrm{BCS}}\approx 0.61\ T_F\ e^{\pi/(2 a k_F)}.
\end{equation}
At zero temperature the expression for the second diagram in
Fig. \ref{fig:lambdaflow}, the particle-hole (ph) loop, vanishes if it is
evaluated for vanishing external momenta, as both poles of the frequency
integration are always either in the upper or lower half of the complex
plane. The dominant part of the scattering in a fermion gas occurs, however,
for momenta on the Fermi surface rather than for zero momentum. For non-zero
momenta of the external particles the particle-hole loop makes an important
contribution.  Setting the external frequencies to zero, we find that the
inverse propagators in the particle-hole loop are
\begin{equation}\label{eq:loopmom1}
P_\psi(q)=i q_0 +(\vec{q}-\vec{p}_1)^2-\mu,\quad \text{and}\quad P_\psi(q)=i q_0 +(\vec{q}-\vec{p}_2^{\,\prime})^2-\mu.
\end{equation}
Depending on the value of the momenta $\vec{p}_1$ and $\vec p_2^{\,\prime}$,
there are now values of the loop momentum $\vec q$ for which the poles of the
frequency integration are in different half-planes so that there is a nonzero
contribution even for $T=0$.

To include the effect of particle-hole fluctuations one could take the full
momentum-dependence of the vertex $\lambda_\psi$ into account. However, the
resulting integro-differential equations represent a substantial numerical
challenge. As a simple and efficient approximation, one therefore restricts the
flow to the running of a single coupling $\lambda_\psi$ by choosing an
appropriate momentum projection.  In the purely fermionic description,
this flow equation has a simple structure and the solution for
$\lambda_{\psi}^{-1}$ can be written as
\begin{equation}\label{PHComp}
 \big(\lambda_{\psi}(k=0)\big)^{-1}=\big(\lambda_{\psi}(k=\Lambda)\big)^{-1} + \mbox{pp-loop} + \mbox{ph-loop}\,.
\end{equation}
Since the ph-loop depends only weakly on the temperature, one can evaluate it
at $T=0$ and add it to the initial value $\lambda_\psi(k=\Lambda)^{-1}$.  As
$T_c$ depends exponentially on the "effective microscopic coupling"
$\left(\lambda_{\psi,\Lambda}^{\text{eff}}\right)^{-1}=(\lambda_{\psi}(k=\Lambda)^{-1}
+ \text{ph-loop})$, any shift in
$\left(\lambda_{\psi,\Lambda}^{\text{eff}}\right)^{-1}$ results in a
multiplicative factor for $T_c$. The numerical value of the ph-loop and
therefore of the correction factor for $T_c/T_F$ depends on the precise
projection description. The averaging prescription used by Gorkov \& Melik-Barkhudarov (1961) leads to
\begin{equation}
T_c=\frac{1}{(4e)^{1/3}} T_{c,\text{BCS}} \approx \frac{1}{2.2} T_{c,\text{BCS}}
\label{eq:}
\end{equation}
and similar for the gap $\Delta$ at zero temperature.


\subsection{Scale-dependent bosonization}\label{sec:Bosonization}

In Sect. \ref{sec:micmodBCSBEC} we describe an effective four-fermion
interaction by the exchange of a boson. In this picture the phase transition
to the superfluid phase is indicated by the vanishing of the bosonic "mass
term" $m^2 = 0$. Negative $m^2$ leads to the spontaneous breaking of
U(1) symmetry, since the minimum of the effective potential occurs for a
nonvanishing superfluid density $\rho_0>0$.
\begin{figure}[b]
\flushleft{\hspace{1.5cm}
\includegraphics[width=0.25\textwidth]{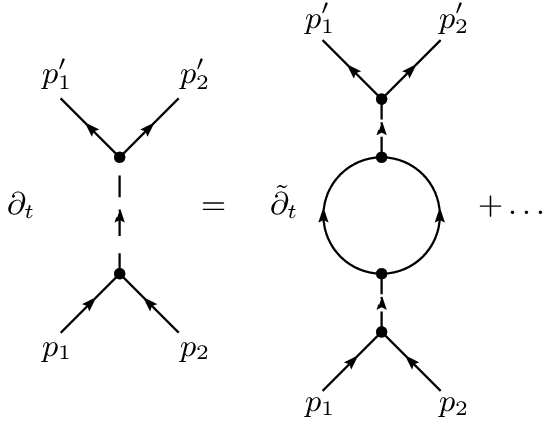}}
\flushright{
\vspace{-3.1cm}
\includegraphics[width=0.35\textwidth]{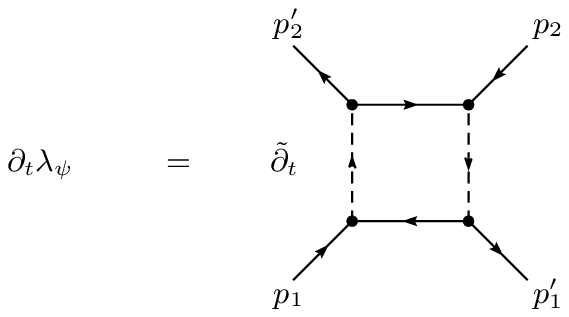}
\hspace{1.5cm}}
\caption{Left panel: Flow of the boson propagator. Right panel: Box diagram for the flow of the four-fermion interaction.}
\label{fig:bosonexchangeloop}
\end{figure}
For $m^2 \geq 0$ we can solve the field equation for the boson $\phi$ as a
functional of $\psi$ and insert the solution into the effective action. This
leads to an effective four-fermion vertex describing the scattering
$\psi_1(p_1)\psi_2(p_2)\to \psi_1(p_1^{\,\prime})\psi_2(p_2^{\,\prime})$
\begin{equation}
\lambda_{\psi,\text{eff}}=\frac{-h^2}{i(p_1+p_2)_0+\frac{1}{2}(\vec p_1+\vec p_2)^2+m^2}.
\label{eq:lambdapsieff}
\end{equation}
To investigate the breaking of U(1) symmetry and the onset of superfluidity,
we first consider the flow of the bosonic propagator, which is mainly driven
by the fermionic loop diagram. For the effective four-fermion interaction this
accounts for the particle-particle loop (see left panel, r.h.s. of
Fig. \ref{fig:bosonexchangeloop}). In the BCS limit of a large microscopic
$m_\Lambda^2$ the running of $m^2$ for $k\to0$ reproduces the BCS result
(Cooper 1956; Bardeen \emph{et al.} 1957).

The particle-hole fluctuations are not accounted for by the renormalization of
the boson propagator. Indeed, we have neglected so far that a four-fermion
interaction term $\lambda_\psi$ in the effective action is generated by the
flow. This holds even if the microscopic pointlike interaction is absorbed by
a Hubbard-Stratonovich transformation into an effective boson exchange such
that $\lambda_\psi(\Lambda)=0$. The strength of the total interaction between
fermions
\begin{equation}\label{eq:lambdapsieff2}
\lambda_{\psi,\text{eff}}=\frac{-h^2}{i(p_1+p_2)_0+\frac{1}{2}(\vec p_1+\vec p_2)^2+m^2} + \lambda_{\psi}
\end{equation}
adds $\lambda_\psi$ to the piece generated by boson exchange. In the partially
bosonized formulation, the flow of $\lambda_\psi$ is generated by the box
diagrams depicted in the right panel of Fig. \ref{fig:bosonexchangeloop}. A
direct connection to the particle-hole diagrams of
Fig. \ref{fig:lambdaflow} can be established on the BCS side 
and in the microscopic regime: There the boson gap $m^2$ is large. In this
case, the effective fermion interaction in Eq. \eqref{eq:lambdapsieff2}
becomes momentum independent, diagrammatically corresponding to a contracted
bosonic propagator.  The box diagram in Fig. \ref{fig:bosonexchangeloop} is
then equivalent to the particle-hole loop investigated in the last section
with the pointlike approximation
$\lambda_{\psi,\text{eff}}\to-\frac{h^2}{m^2}$ for the fermion interaction
vertex.

In contrast to the particle-particle fluctuations (leading to SSB for
decreasing $T$), the particle-hole fluctuations lead only to quantitative
corrections and depend only weakly on temperature. This can be checked
explicitly in the pointlike approximation, and holds not only in the BCS
regime where $T/\mu \ll 1$, but also for moderate $T/\mu$ as realized at the
critical temperature in the unitary regime. We therefore evaluate the
box diagrams in Fig. \ref{fig:lambdaflow} for zero temperature. We emphasize that
a temperature dependence, resulting from the couplings
parameterizing the boson propagator, is implicitly taken into account. For the external momenta we use an averaging on the Fermi surface similar to the one of Gorkov \& Melik-Barkhudarov (1961). For details see Floerchinger \emph{et al.} (2009).

After these preliminaries, we can now incorporate the effect of particle-hole
fluctuations in the RG flow.  In principle one could simply take $\lambda_\psi$ as an additional coupling into account. However, it is much more elegant to use a scale-dependent Hubbard-Stratonovich transformation (Gies \& Wetterich 2001; Pawlowski 2007; Floerchinger \& Wetterich 2009) which absorbes $\lambda_\psi$ into the Yukawa-type interaction with the bosons at every scale $k$. By construction, there is then no self interaction between the fermionic quasiparticles. The general procedure of ``partial bosonization'' is discussed in detail in Floerchinger \emph{et al.} (2009). A slightly modified scheme based on the exact flow equation derived in (Floerchinger \& Wetterich 2009) has been used in (Floerchinger \emph{et al.} 2010).  In that formulation one finds for the renormalized coupling $m^2$ in the symmetric regime an additional term reflecting the absorption of $\lambda_\psi$ into the fermionic interaction induced by boson exchange,
\begin{equation}
\partial_t m^2 = \partial_t m^2{\big |}_\text{HS} + \frac{m^4}{h^2} \partial_t \lambda_\psi {\big |}_\text{HS}.
\label{eq:modifiedflowminSYM}
\end{equation}
Here $\partial_t m^2{\big |}_\text{HS}$ and $\partial_t \lambda_\psi {\big |}_\text{HS}$ denote the flow equations when the Hubbard-Stratonovich transformation is kept fixed. Since $\lambda_\psi$ remains now zero during the flow, the effective four-fermion interaction $\lambda_{\psi,\text{eff}}$ is purely given by the boson
exchange. However, the contribution of the particle-hole exchange diagrams is incorporated via the second term in Eq. \eqref{eq:modifiedflowminSYM}.
The flow equation of all other couplings are the same as with fixed Hubbard-Stratonovich transformation. In the regime with spontaneous symmetry breaking we use
\begin{eqnarray}
\nonumber
\partial_t h &=& \partial_t h {\big |}_\text{HS} + \frac{\lambda \rho_0}{h} \partial_t \lambda_\psi {\big |}_\text{HS},\\
\nonumber
\partial_t \rho_0 &=& \partial_t \rho_0{\big |}_\text{HS} - 2 \frac{\lambda \rho_0^2}{h^2} \partial_t \lambda_\psi{\big |}_\text{HS},\\
\partial_t \lambda &=& \partial_t \lambda{\big |}_\text{HS} + 2 \frac{\lambda^2 \rho_0}{h^2} \partial_t \lambda_\psi{\big |}_\text{HS}.
\end{eqnarray}

We emphasize that our non-perturbative flow equations go beyond the treatment
by Gorkov \& Melik-Barkhudarov (1961) which includes the particle-hole
diagrams only in a perturbative way. Furthermore, the inner bosonic lines
$h^2/P_\phi(q)$ in the box diagrams include the center-of-mass momentum
dependence of the four-fermion vertex. This is neglected in Gorkov's pointlike
treatment, and thus represents a further improvement of the classic
calculation. Actually, this momentum dependence becomes substantial away from
the BCS regime where the physics of the bosonic bound state sets in. The continuous description of dynamically
transmuting degrees of freedom is a particular strength of an RG description,
as exemplified also in the context of QCD (Gies \& Wetterich 2004,
Braun 2009).

\section{Running Fermion sector}\label{sec:BCSBECrunfermion}

In this section we aim at a systematic extension of the truncation scheme and
consider a running fermion sector. Similar parameterizations of the fermionic
self-energy have been studied in Gubbels \& Stoof (2008), Bartosch \emph{et al.} (2009) and Strack \emph{et al.} (2008). Further, we include an atom-dimer interaction term. This section is based on the work by Floerchinger \emph{et al.} (2010).


\subsection{Completion of the truncation}
\label{sec:floweq}

In addition to the running couplings that have occurred so far in
Secs. \ref{sec:BCSBECtrunc}, now we want to take into account $k$-dependent
parameters $\bar m_\psi^2$ and $Z_\psi$, in order to parametrize
fluctuation effects on the self-energy of the fermionic quasiparticles. 
The extension of the truncation explicitely reads
\begin{eqnarray}
\nonumber
\Gamma_k & =  & \int_0^{1/T}d\tau \int d^3x {\bigg \{} \bar\psi^\dagger Z_\psi(\partial_\tau -\Delta) \bar\psi +\bar m_\psi^2\bar\psi^\dagger\bar\psi+\bar{\phi}^*\left(\bar Z_\phi \partial_\tau-\frac{1}{2}A_\phi \Delta\right)\bar\phi  \\
&{}&\quad\quad\quad\quad\quad\quad + \bar U(\bar \rho,\mu) -\bar h (\bar \phi^* \bar\psi_1\bar\psi_2 + \bar \phi \bar\psi_2^\ast \bar\psi_1^\ast )+\bar\lambda_{\phi\psi}\bar\phi^\ast\bar\phi\bar\psi^\dagger\bar\psi {\bigg \} }.
\label{eq:baretruncationFERM}
\end{eqnarray}
The additional inclusion of the atom-dimer coupling $\bar \lambda_{\phi\psi}$ closes the
truncation on the level of interaction terms quartic in the fields
and describes three-body scattering (Diehl \emph{et al.} 2007$c$). It leads to quantitative modifications for the many-body problem. In the regime of spontaneously broken symmetry ($\rho_0>0$) the atom dimer coupling
leads to a modification of the Fermi surface, in addition to the gap
$\sqrt{h^2 \rho_0}$. 

We define the renormalized fields $\phi=A_\phi^{1/2}\bar\phi$,
$\rho=A_\phi \bar \rho$, $\psi=Z_\psi^{1/2}\bar\psi$ and study the flow
  of the renormalized couplings $Z_\phi=\bar Z_\phi / A_\phi$, $h=\bar h/(A_\phi^{1/2} Z_\psi)$,
$\lambda_{\phi\psi}=\bar \lambda_{\phi\psi}/(A_\phi Z_\psi)$, $m_\psi^2=\bar
m_\psi^2/Z_\psi$.
%
%
As before, we expand the effective potential in monomials of $\rho$, see
\Eqref{eq:BCSBECueff}. 
We use again a purely space-like regulator which is
adjusted to the running Fermi surface,
\begin{equation}
\nonumber
R_{k,\psi}=Z_\psi\big[\text{sign}(\vec p^2-r_F^2)k^2- (\vec p^2-r_F^2)\big)]\theta\left(k^2-|\vec p^2-r_F^2|\right), \quad
R_{k,\phi}=A_\phi \left[k^2-\vec p^2/2\right]\theta\left(k^2-\vec p^2/2\right),
\label{eq:cutoffFERM}
\end{equation}
where $r_F^2= - m_\psi^2-\lambda_{\phi\psi}\rho_0$. For the fermions it
regularizes fluctuations around the running Fermi surface, while for the
bosons fluctuations with small momenta are suppressed.

\begin{figure}[t!]
\centering
\includegraphics[scale=0.65]{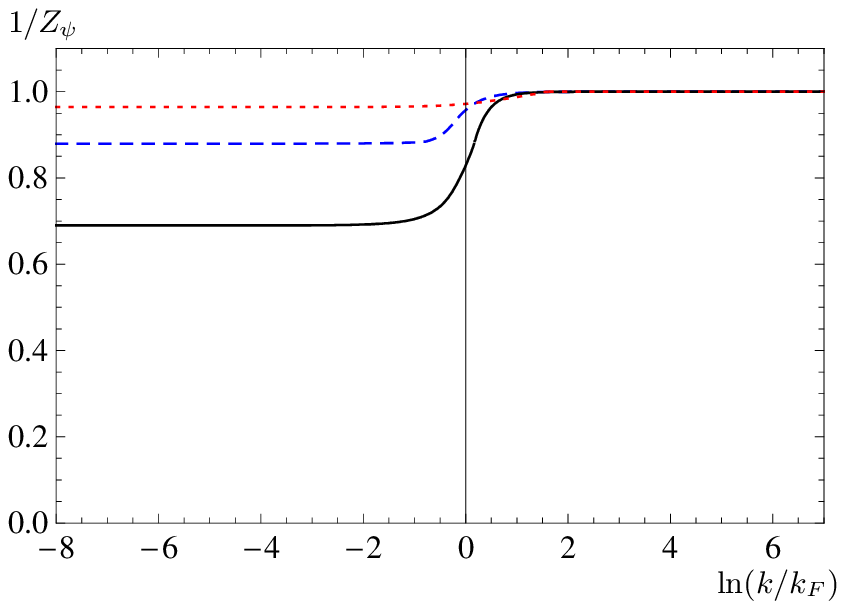}\hspace{0.8cm}
\includegraphics[scale=0.65]{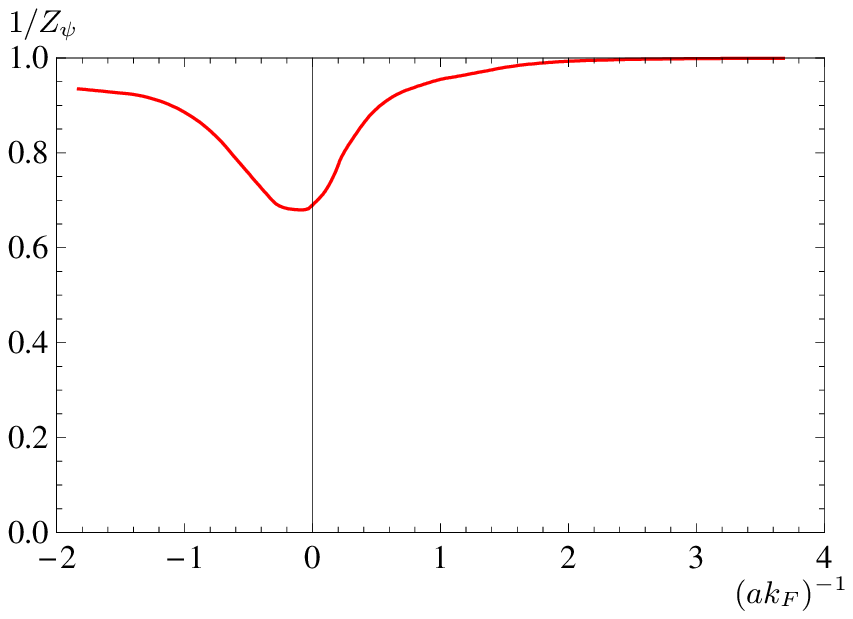}
\caption{Left panel: Flow of the inverse fermionic wave function renormalization
  $1/Z_\psi$ at $T=0$ at three different points of the crossover: $c^{-1}=-1$ (dashed line), $c^{-1}=0$ (solid line), $c^{-1}=1$ (short dashed line). Right panel: Inverse fermionic
  wave function renormalization $1/Z_\psi$ at $k=0$ as a
  function of the crossover parameter $c^{-1}$.}\label{fig:zpsiflow}
\end{figure}

As a first quantity we investigate the vacuum dimer-dimer scattering length $a_M$ expressed in units of the
atom-atom scattering length $a$. On the BEC side of the resonance we can
derive this quantity from the corresponding couplings by the equation
\begin{equation}
\frac{a_M}{a}=2\frac{\lambda}{\lambda_{\psi,\mathrm{eff}}},\quad \lambda_{\psi,\mathrm{eff}}=8\pi a=\frac{8\pi}{\sqrt{-\mu}},\ \text{for}\ \mu <0\ \text{and}\ k=0.
\end{equation}
To explicitely compute the vacuum quantity $a_M/a$, we choose a value
  for $a$ on the far BEC side, where for broad resonances the identity
  $a=(-\mu)^{-1/2}$ holds. We evolve the flow of the couplings to the infrared
  and extract the value of $\lambda$, completely fixing $a_M/a$. In this
truncation including the $\lambda_{\phi\psi}$ vertex, we find $a_M/a=0.59$, which is in very good
agreement with the well-known result from a direct solution of the
Schr\"odinger equation, $a_M/a=0.60$ (Petrov \emph{et al} 2004). The accuracy
of this result is somewhat surprising since no momentum dependence of
$\lambda_{\phi\psi}$ has been taken into account. The latter has turned out to
be important for the atom-dimer scattering Diehl \emph{et al.} (2007$c$).  On
the other hand, the general leading-order effect of fermionic
momentum-dependencies is captured by the wave function renormalization
$Z_\psi$ the effect of which is largest in the strongly interacting regime,
when $|c^{-1}|<1$, cf. Fig. \ref{fig:zpsiflow}.
%

\subsection{Fermi sphere and dispersion relation}
The dispersion relation can be computed from the determinant of the renormalized fermionic propagator 
\begin{equation}
G_\psi^{-1} = \begin{pmatrix} -h \phi_0 \epsilon&-\omega -(\vec q^2+m_\psi^2+\lambda_{\phi\psi} \rho_0) \\ 
-\omega + (\vec q^2+m_\psi^2+\lambda_{\phi\psi} \rho_0)& h \phi_0 \epsilon \end{pmatrix}.
\end{equation}
by the equation $\text{det}\,G_\psi^{-1}=0$. Here, we have evaluated $G_\psi^{-1}$ in the regime of  spontaneously broken symmetry and performed analytical continuation to real frequencies $\omega$. We see that the dispersion relation is affected by the running of the couplings $Z_\psi$, $m_\psi^2$ and $\lambda_{\phi\psi}$, which follows as $\omega =
\pm \sqrt{\Delta^2+(\vec q^2-r_F^2)^2}$, where $\Delta=h\sqrt{\rho_0}$ is the
gap and $r_F=\sqrt{-m_\psi^2-\lambda_{\phi\psi}\rho_0}$ is the effective
radius of the Fermi sphere.
\begin{figure}[t!]
\centering
\includegraphics[scale=0.65]{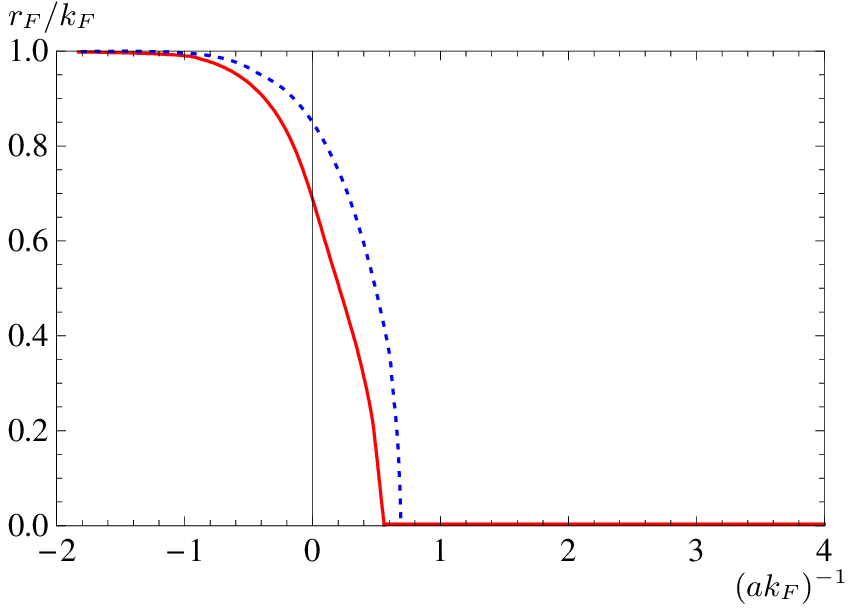}\hspace{0.8cm}
\includegraphics[scale=0.65]{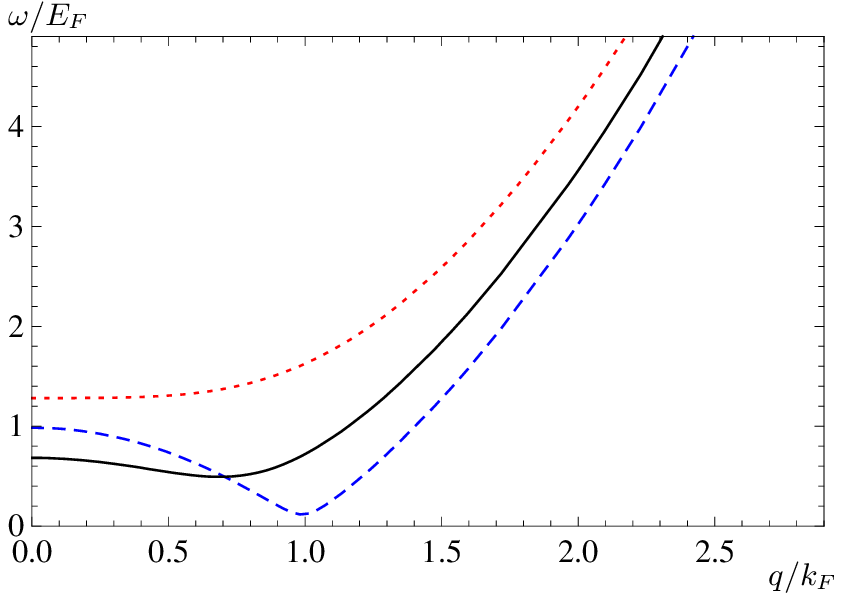}
\caption{Left panel: Effective Fermi radius $r_F/k_F$ as a function of the
  crossover parameter $c^{-1}$ for vanishing temperature (solid line). We compare to the effective Fermi radius in an approximation
  without the contribution of the atom-dimer vertex
  $\lambda_{\phi\psi}$ (dotted line). Right panel: Positive branch of the
  dispersion relation $\omega(q)$ in units of $E_F$  for $c^{-1}=-1$
  (dashed), $c^{-1}=0$ (solid) and $c^{-1}=1$ (short dashed).}
\label{fig:fermiradius}
\end{figure}
On the far BCS side of the crossover the
renormalization effects on $r_F$ are small and $r_F$ approaches
its classical value
$r_F\simeq\sqrt{\mu}=k_F$ where $k_F=(3\pi^2 n)^{1/3}$. Close to the resonance,
the Fermi sphere gets smaller. It finally vanishes on the BEC side at a point with
$c^{-1}\approx 0.6$, see Fig. \ref{fig:fermiradius}. Here, the fermions are gapped even for
$\Delta\to0$ by a positive value of $m_\psi^2+\lambda_{\phi\psi} \rho_0$.
%

\section{Results}

For the studies in section \ref{sec:BCSBECrunfermion}, we have omitted the effect of
particle-hole fluctuations for simplicity. In the following, however, all the
results are given for the correspondingly extended truncation including
particle-hole fluctuations.

\subsection{Single-particle gap at $T=0$}

As a first study including all the couplings introduced in this contribution we investigate the single particle gap at zero temperature. 
On the far BCS side it is possible to compare to the result by Gorkov \& Melik-Bakhudarov (1961) which is given by $\Delta/E_F=
(2/e)^{7/3}e^{\pi/(2c)}\label{eq:gapgorkov}$. Our approach allows to extend to the strongly interacting regime and even to the BEC side of the
crossover, see Fig.\ \ref{fig:tcrit}. At the unitary point, $(ak_F)^{-1}=0$, we
obtain $\Delta/E_F=0.46$. Further, we compare our result for chemical potential in units of the Fermi energy at the unitary point $\mu/E_F=0.51$ to different (non-perturbative) methods in
Tab. \ref{tab:fpvaluesSSB}.
\begin{table}
\footnotesize
\centering
\begin{tabular}{ccc}\hline
&$\mu/E_F$ & $\Delta/E_F$ \\ \hline
Carlson \emph{et al.} (2003) (QMC) & 0.43 & 0.54 \\
Perali \emph{et al.} (2004) (t-matrix approach) & 0.46 & 0.53\\
Haussmann \emph{et al.} (2007) (2PI)& 0.36 & 0.46 \\
Bartosch  \emph{et al.}  (2009) (FRG, vertex exp.)& 0.32 & 0.61 \\
Floerchinger  \emph{et al.}  (2010) (FRG, derivative exp.) & 0.51 & 0.46 \\ \hline
\end{tabular}
\caption{\label{tab:fpvaluesSSB}Results for the single-particle gap and the
  chemical potential at $T=0$ and at the unitary point by various authors.} 
\end{table}

\subsection{Phase diagram}\label{crittemp}

\begin{figure}
\centering
\includegraphics[width=0.45\textwidth]{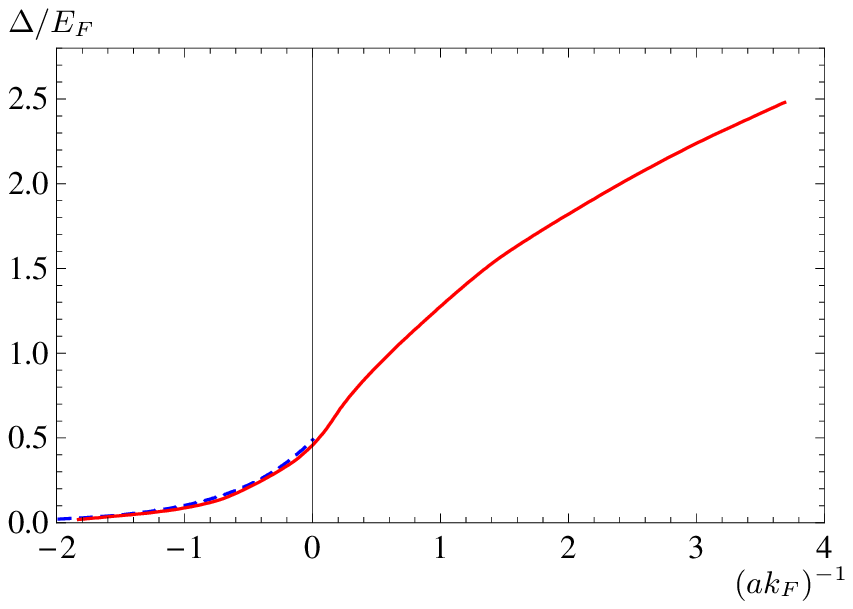}\hspace{0.8cm}
\includegraphics[width=0.48\textwidth]{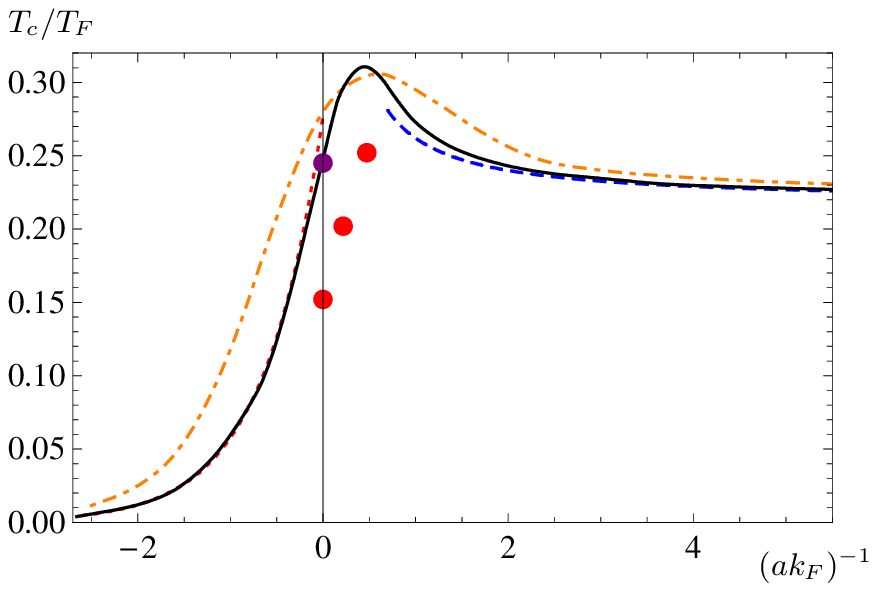}
\caption{Left panel: Gap in units of the Fermi energy $\Delta/E_F$ as a
  function of $(ak_F)^{-1}$ (solid line). For comparison, we also plot the
  result found by Gorkov and Melik-Bakhudarov (dashed) and extrapolate
  this to the unitary point $(ak_F)^{-1}=0$, where
  $\Delta_{\mathrm{GMB}}/E_F=0.49$. 
	Right panel: Critical temperature $T_c/T_F$ in units of the Fermi temperature as a function of the crossover
  parameter $(ak_F)^{-1}$. The solid line gives the result of our full computation refered to in the text. The dot-dashed line is obtained using the more basic truncation discussed in section \ref{sec:BCSBECtrunc}. The dotted line shown for $(ak_F)^{-1}<0$ shows the result of the perturbative calculation by Gorkov \& Melik-Barkhudarov (1961). The dashed line corresponds to an interacting BEC with the shift in $T_c$ according to Eq.\ \eqref{eq:shiftTcBEC}. We use here $a_\text{M}/a=0.60$ and $\kappa=1.31$. The three red dots close to and at unitarity show
  the QMC results by Burovski \emph{et al.} (2008), while the single purple dot gives the result of Akkineni \emph{et al.} (2007).}\label{fig:tcrit}
\end{figure}

Our results for the critical temperature for the phase
transition to superfluidity throughout the crossover are shown in
Fig. \ref{fig:tcrit}. We plot the critical temperature in units of the Fermi
temperature $T_c/T_F$ as a function of the scattering length measured in units
of the inverse Fermi momentum, i.~e. the concentration $c=a k_F$.

On the BCS side of the crossover, where $c^{-1}\ll-1$, the BCS approximation and the effect of particle-hole
fluctuations yield a critical temperature (Gorkov \& Melik-Barkhudarov 1961)
\begin{equation}
\frac{T_c}{T_F}=\frac{e^C}{\pi}\left(\frac{2}{e}\right)^{7/3} e^{\pi/(a k_F)}\approx 0.28 e^{\pi/(a k_F)}.
\end{equation}
depicted by the short dashed line in the right panel of
Fig. \ref{fig:tcrit}. Here, $C \approx 0.577$ is Euler's constant.  On the
BEC side for very large and positive $c^{-1}$, our result approaches the
critical temperature of a free Bose gas where the bosons have twice the mass
of the fermions $M_B=2M$. In our units the critical temperature is then
\begin{equation}
\frac{T_{c,\text{BEC}}}{T_F}=\frac{2\pi}{\left(6\pi^2 \zeta(3/2)\right)^{2/3}}\approx 0.218.
\end{equation} 
In-between there is the unitarity regime, where the two-atom scattering length
diverges ($c^{-1} \rightarrow 0$) and we deal with a system of strongly
interacting fermions.

Our best result including particle-hole fluctuations is given by the
solid line. This may be compared with a functional renormalization flow
investigation ignoring particle-hole fluctuations as discussed in section \ref{sec:BCSBECtrunc} (dot-dashed line)
(Diehl \emph{et al.} 2007$a$). For $c\to 0_-$ the solid line of our result
agrees with
BCS theory including the correction by Gorkov \&
Melik-Barkhudarov (1961). Deviations from this perturbative regime appear only rather close to the regime of strong interactions $c^{-1}\rightarrow 0$.

For $c\to 0_+$ this value is approached in the form (Baym \emph{et al.} 1999)
\begin{equation}
\frac{T_c-T_{c,\text{BEC}}}{T_{c,\text{BEC}}}=\kappa a_M n_M^{1/3}=\kappa \frac{a_M}{a}\frac{c}{(6\pi^2)^{1/3}}.
\label{eq:shiftTcBEC}
\end{equation}
Here, $n_M=n/2$ is the density of molecules and $a_M$ is the molecular
scattering length. Using our result $a_M/a=0.59$ obtained from solving the flow
equations in vacuum, the coefficient determining the shift in $T_c$ compared to the
free Bose gas yields $\kappa=1.39$,  see also Diehl \emph{et al.}
(2007$a$). In (Arnold \& Moore 2001;
  Kashurnikov \emph{et al.} 2001), the result
  for an interacting BEC is determined as
  $\kappa = 1.31$ (dashed curve on BEC side of \ref{fig:tcrit}), see also
  (Blaizot \emph{et al.} 2006$a,b$) for a functional RG study. This is
in reasonable agreement with our result. As further characteristic quantities we give
the maximum of the ratio $(T_c/T_F)_\text{max}\approx0.31$ and the location of
the maximum $(ak_F)^{-1}_\text{max}\approx0.40$.  For $c^{-1}>0.5$, the effect
of the particle-hole fluctuations vanishes. This is expected, since the
chemical potential is now negative $\mu<0$ such that the  Fermi surface disappears.

In the unitary regime ($c^{-1}\approx 0$), the particle-hole fluctuations still
have a quantitative effect. We can give an improved estimate for the critical
temperature at the resonance ($c^{-1}=0$) where we find $T_c/T_F=0.248$ and a
chemical potential $\mu_c/T_F=0.55$. A comparison to other methods and our
previous work is given in Tab. \ref{tab:tcritunitarity}. We observe reasonable
agreement with QMC results for the chemical potential $\mu_c/T_F$, however,
our critical temperature $T_c/T_F$ is larger.

\begin{table}
\footnotesize
\centering
\begin{tabular}{ccc}\hline
&$\mu_c/E_F$ & $T_c/T_F$ \\ \hline
Burovski \emph{et al.} (2006) (QMC) & 0.49 & 0.15 \\
Bulgac \emph{et al.} (2006) (QMC) & 0.43 & $<$ 0.15\\
Akkineni \emph{et al.} (2007) (QMC) &- & 0.245 \\
Previous FRG estimate by Floerchinger \emph{et al.} (2009) & 0.68&  0.276 \\
Floerchinger  \emph{et al.}  (2010) (FRG) & 0.55 & 0.248 \\ \hline
\end{tabular}
\caption{\label{tab:tcritunitarity}Results for $T_c/T_F$ and $\mu_c/T_F$ at
  the unitary point by various authors.} 
\end{table}


\section{Discussion and Outlook}
\label{sec:conc}

As illustrated with the example of the BCS-BEC crossover, the functional
  RG is capable of describing strongly-interacting many-body systems in a
  consistent and controllable fashion. Once the relevant degrees of freedom
  are identified -- possibly in a scale-dependent manner -- approximation
  schemes based on expansions of the effective action can be devised that
  facilitate systematically improvable quantitative estimates of physical
  observables. For the BCS-BEC crossover, already a simple derivative
  expansion including fermionic and composite bosonic degrees of freedom
  exhibits all qualitative features of the phase diagram.

The inclusion of particle-hole fluctuations and higher-orders of the
  derivative expansion improve our numerical results in the BCS as well as in the BEC limit
  of the crossover in agreement with well-known other
field-theoretical methods. We obtain satisfactory quantitative
precision on the BCS and BEC sides of the resonance. Remarkably, the
  functional RG allows for a description of both, many-body as well as
  few-body physics within the same formalism. For instance, our result for the
  molecular scattering length ratio $a_M/a$ is in good agreement with the
exact result (Petrov \emph{et al.} 2004). This quantitative accuracy is
remarkable, as we have started with a purely fermionic microscopic theory
without propagating bosonic degrees of freedom or bosonic interactions.

In the strongly interacting regime where the scattering length diverges, no
exact analytical treatments are available. Our results for the gap
$\Delta/E_F$ and the chemical potential $\mu/E_F$ at zero temperature are in
reasonable agreement with Monte-Carlo simulations. This holds also for the
ratio $\mu_c/E_F$ at the critical temperature. The critical temperature
$T_c/T_F$ itself is found to be larger than the Monte-Carlo result.

In future studies, our approximations might be improved mainly at two
points. One is the frequency- and momentum dependence of the boson
propagator. In the strongly interacting regime, this might be rather involved,
developing structures beyond our current approximation. A more detailed
resolution might lead to modifications in the contributions from bosonic
fluctuations to various flow equations. Another point concerns structures in
the fermion-fermion interaction that go beyond a diatom bound-state exchange
process. Close to the unitary point, other contributions might arise, for
example in form of a ferromagnetic channel. While further quantitative
modifications in the unitarity regime are conceivable, the present 
approximation already allows for a coherent description of the BCS-BEC
crossover for all values of the scattering length, temperature and density by
one simple method and microscopic model. This includes the critical behavior
of a second order phase transition as well as the vacuum, BEC and BCS limits.

In order to improve the comparison between the QMC simulations and the
  functional RG results in the strongly interacting regime, the Wetterich
  equation can also be evaluated in a finite volume. This may shed light on
  possible finite size effects in the QMC simulations, and can help to
  quantitatively compare finite-volume studies with infinite-volume
  calculations inherent to most analytical works.

\begin{acknowledgements}
  The authors are grateful to J.~Braun, S.~Diehl, J.~M.~Pawlowski, C.~Wetterich for collaboration on the subject reviewed here.
  This work has been supported by the DFG research unit FOR
  723. H.G. acknowledges support by the DFG under contract Gi 328/5-1
  (Heisenberg program). S.F. acknowledges support by the Helmholtz Alliance HA216/EMMI.
\end{acknowledgements}


\end{document}